\documentclass[journal,comsoc]{IEEEtran}
\usepackage[T1]{fontenc}
\usepackage[numbers,sort&compress]{natbib}
\usepackage[pdftex]{graphicx}
\usepackage{epstopdf}
\usepackage{amsmath}
\usepackage{algorithm}
\usepackage{algpseudocode}
\usepackage{booktabs}
\usepackage[table, x11names]{xcolor}
\usepackage[caption=false,font=footnotesize]{subfig}
\usepackage[inline]{enumitem}
\usepackage{verbatim}
\usepackage{etoolbox}
\makeatletter
\patchcmd{\@makecaption}
  {\scshape}
  {}
  {}
  {}
\makeatother

\newcommand{\datM}{eMail}
\newcommand{\datE}{Epinions}
\newcommand{\datW}{Wiki-Vote}
\newcommand{\datD}{DBLP}
\newcommand{\datC}{CA-Cond-Mat}
\newcommand{\algD}{degree}
\newcommand{\algE}{eigen vector}
\newcommand{\algK}{$k$-shell}
\newcommand{\algP}{Ks-P}
\newcommand{\algH}{Ks-Hp}
\newcommand{\algX}{Pr}

\usepackage[iso]{datetime}

\newcommand{\reffig}[1]{Fig.~\ref{#1}}

\newcommand{\reftbl}[1]{Table~\ref{#1}}

\usepackage{amsmath, amssymb,amsfonts,amsthm}

\newcommand{\hbAbs}[1]{| #1 |}	
\theoremstyle{plain}

\theoremstyle{definition}

\theoremstyle{remark}

\newcommand{\hSoR}  {\ensuremath{\mathbb{R}}}      
\begin{document}
\title{
	Computational Models for Commercial Advertisements in Social Networks
}
\author{
	Samet~Atdag and Haluk~O.~Bingol%
	\thanks{
	S.~Atdag and H.~O.~Bingol are with 
	the Department of Computer Engineering, 
	Bogazici University, 
	Istanbul,
	34342 Turkey 
	e-mail: samet.atdag@boun.edu.tr.
	}%
}

\maketitle

\begin{abstract}
            Identifying noteworthy spreaders in 
            a network is essential for understanding 
            the spreading process and controlling the 
            reach of the spread in the network. 
            The nodes that are holding more intrinsic 
            power to extend the reach of the spread are 
            important due to demand for various 
            applications such as viral marketing, 
            controlling rumor spreading or get a 
            better understanding of spreading of 
            the diseases. 
            As an application of the viral marketing, 
            maximization of the reach with a fixed 
            budget is a fundamental requirement in 
            the advertising business.
            Distributing a fixed number of promotional 
            items for maximizing the viral reach  
            can leverage influencer detection methods.
            For detecting such ``influencer'' nodes, 
            there are local metrics such as degree 
            centrality (mostly used as in-degree 
            centrality) or global metrics such as \algK\ 
            decomposition or eigenvector centrality. 
            All the methods can rank graphs but they 
            all have limitations and there is still 
            no de-facto method for influencer detection 
            in the domain. 

            In this paper, we propose an \textbf{extended 
            \algK\ algorithm} which better utilizes 
            the \algK\ decomposition for identifying 
            viral spreader nodes using the topological 
            features of the network.
            We use Susceptible-Infected-Recovered model 
            for the simulations of the spreading
            process in real-life networks and the 
            simulations demonstrates that our approach 
            can reach to up to 36\% larger crowds within 
            the same network, with the same number of 
            initial spreaders.
\end{abstract}


\IEEEpeerreviewmaketitle

\section{Introduction}

\IEEEPARstart{I}{n}  
our economy driven society, competition is the 
hidden force triggering advertisements. 
Companies race for selling functionally similar 
(if not same) products, desiring to reach out each 
single buyer and advertise their inventory. 
Conventional commercial advertisement methods 
generally consist some kind of broadcasting such 
as ads in newspapers, magazines, radio and TV 
commercials.

Broadcasting has an advantage of reaching masses 
with a single advertisement, however, 
it bears several disadvantages. 
First of all, there is no targeting. 
The advertisement reaches to all, even they are 
not in the target group, which is usually very 
expensive and it exceeds reasonable budgets. 
Secondly, since there is no controlling authority, 
competitors tend to overrating (or even faking) 
their content. This causes a loss in trust and 
this leads to an increase in importance of 
viral marketing: in response to the information 
pollution of commercials, consumers (or buyers) 
have lost their trust throughout the advertisement 
exposition and people start to depend on either 
their firsthand experience for similar products 
or they depend on experiences of other people 
they trust.

Word-of-mouth diffusion is a well-known phenomena 
by which information can reach to large populations, 
possibly influencing public opinion, causing 
adoption of innovations or brand awareness~\cite{%
	bakshy2011everyone}. 
In a network of people, there are special 
individuals whom people have a certain trust level. 
These special people demonstrate desirable 
attributes such as credibility, expertise, or 
enthusiasm, or the connections they have in the 
network allow them to influence a 
disproportionately large number of others, 
directly or indirectly via a cascade of the 
influence~\cite{%
	glad00}.
These special individuals are called as 
\emph{influencers}. 
What those special individuals say has a potential 
to influence the people on their adjacency and may 
cause a word-of-mouth diffusion effect.

Detection of influencers in a network is a hot 
topic and there are numerous methods proposed for 
achieving the same goal. 
Since networks get larger in scale, finding a 
single influencer can not be sufficient to diffuse 
the desired information throughout a large portion 
of the network. 
Instead of starting with only one influencer, it 
is an intuitive action to choose a set of 
influencers and push the desired information via 
the set of influencers and maximizing the reach of 
the information in the network. 
The proportion of the individuals in the network 
who are directly or indirectly (via a cascading 
effect) informed about the information to the 
whole network is called \emph{network coverage} 
of the set of influencers. 
Now the problem is finding a minimum set of 
influencers to maximize the network coverage.

Instead of finding all influencers in a 
network, which may become impractical even in 
small networks, if we can find a smart way of 
delivering a limited number of promotional items 
to the most meaningful people in the network. 
Then it may become a way of imposing hidden 
advertisements with much lower costs. 
A simple approach is using viral marketing 
strategy, which depends on finding the most 
influencing set of people in the community and 
delivering a limited number of promotional items 
among the influencers. 
After these promotional deliveries, we expect that 
the piece of information reaches to the majority 
of the social network.

\section{Background}

There are several mathematical and computational 
models to represent networks and influencers~\cite{
	anderson1992infectious,%
	raghavan2007near,%
	daley1965stochastic,%
	kitsak2010identification,%
	lu2011leaders,%
	zhang2016identifying,%
	hajian2011modelling%
}. 

It is a common approach to label each node with a 
(preferably numeric) metric, then sorting all 
the nodes according to the given numeric label, 
and consider the top ones.
One assumption is that an influencer should be 
highly connected.
Then, if all nodes have the same importance,
``degree'' is the most intuitive network metric 
used for this goal.
We can extend degree centrality to the cases 
where nodes have different importance.
If being connected to important nodes become 
important then we end up with ``eigenvalue 
centrality''.
``\algK\ decomposition'' model is another method 
for dismantling the social network 
and grouping similar individuals together 
according to their topological attributes 
in the network~\cite{%
	kitsak2010identification}.

There are many more other methods such as 
LeaderRank, VoteRank, InfluenceRank, etc.~\cite{
	lu2011leaders,%
	zhang2016identifying,%
	hajian2011modelling%
} 
mostly specialized with a single type of 
network, such as Twitter social network. 
Since we try to find a generalized way, 
we did not use the mentioned specialized models.

\subsection{Definitions and Approach}

Let $G(V,E)$ be a network,
where 
$V = \{ 1, \dotsc, N \}$ is the set of vertices 
and $E$ is the set of edges. 
Let $N = \hbAbs{V}$ and $M = \hbAbs{E}$. 
Let $k_i$ be the degree of node $i$ in the network. 

Let function $f_{X} \colon V \to \hSoR$ 
assign a number to each node according 
to criterion $X$.
Suppose the number assigned to a node is 
related to its influence
such as degree, eigenvalue and \algK\ 
number.
Then we rank the nodes according to their 
influence values and select the top $n$ 
as the most influential $n$ nodes.

\subsection{Metrics}

An influencer is opinionated, respected and well-connected. 
In a social network, individuals who have 
connections to many others might have 
more influence, 
more access to information 
or more prestige 
than those who have less connections~\cite{%
	newman2018networks}.
There are influencers having a hybrid combination 
of these three attributes, such as a 
well-connected and opinionated individual may earn 
respect in the network and become an influencer. 
Since well-connectedness is easy to compute,
there are several metrics on it. 
In network science, 
well-connectedness is commonly associated with 
centrality.
Below given the three centrality metrics 
we used for benchmarking.

\begin{itemize}

	\item 
    	\textit{Degree centrality} (Dg). 
	Degree centrality measures connectivity 
	when every node is of the same importance~\cite{%
		newman2018networks}.
	\emph{Degree} $k_{i}$ of a node $i$ is defined 
	as the number of nodes that it is directly connected to.
	Then define number assignment as $f_{Dg} (i) = k_{i}$.

	\item 
	\textit{Eigenvector centrality} (Eg). 
	If we extend degree centrality to the cases 
	where connected to an important node is more 
	important, we end up with eigenvalue centrality~\cite{%
		newman2018networks}.
	
	\item 
	\textit{PageRank} (Pr). 
	Another node ranking commonly used is PageRanking~\cite{%
		newman2018networks}.
\end{itemize}

\subsection{\algK\ Decomposition}

\algK\ decomposition method partitions the network 
into a layered structure, called \emph{shell}, 
which is similar to the structure of an onion. 
The innermost shell is called \emph{core} 
and other shells are called 
\emph{$k$-shells}~\cite{%
	garas2012k}.

This way, 
the method assigns an integer \emph{shell index} $s_{i}$ to each node $i$ 
that represents the connectivity patterns of the node in the network. 

Nodes located at the periphery of the network
will have low values of $s$, while nodes 
located at the inner shells are assigned 
higher values of $s$.

\algK\ decomposition method prunes the network 
iteratively and in each iteration, it removes 
nodes according to their degrees. 

The pruning process first removes the nodes of degree 1,
i.e., $k_{i} = 1$. 
As a result, some nodes with $k_{i} =2$ become of degree 1.
Pruning repeats until there is no node with $k_{i} = 1$. 
Hence the remaining nodes have degree $k_{i} \ge 2$.
Nodes pruned at this stage are labeled with shell index of 1
i.e., $s_{i} = 1$.
The next stage is pruning of nodes of degree $k_{i} = 2$,
which results of nodes of shell index of 2. 
Pruning repeats the process on the remaining network 
for higher $k_{i}$ to extract other shells. 
The process runs until there is no node left. 
At the end, 
every node $i$ is labeled with its corresponding 
shell index $s_{i}$.
Note that all the nodes in the same shell share 
the same shell index.
So shell index by itself is not a very refined 
ranking since many nodes will be ranked the same.

Kitsak et al.~\cite{kitsak2010identification} states 
that the most strongly connected nodes 
who may have the strongest probability of 
spreading occupy high $k$-shells. 
Since nodes laying in same $k$-shell 
roughly have similar connectivity, 
they perform similar spreading capabilities.

\subsection{Infection Spreading Model}

Ideas spread like infections spread, 
using the connections between people 
in a network of people.
We used infection spreading model to 
compare our approach to other approaches.

A very common way to model the infection 
spreading mechanics is following compartmental 
models in epidemiology. 
In these type of models, the population is 
divided into compartments and it is assumed 
that the individuals in same compartment 
perform same characteristics. 
Infection Spreading Model~\cite{%
	anderson1992infectious} 
is a popular yet simple compartmental model, 
which is able to represent the spreading process. 
There are three compartments in foundation: 
Susceptible (S), 
Infected (I) and 
Recovered (R). 
In principle, 
a susceptible node becomes infected 
as a result of getting in touch with an infected node. 
Various models describe what happens after a node 
gets infected.

In SIS model, a susceptible node may become 
susceptible after they got infected. 
In SIR, a susceptible node may recover after 
infection. 
In SI model, an individual lives with the 
infection till they die.
The models have two parameters. 

Parameter $\beta$ is the probability of a 
susceptible contracting the infection 
in the case of interacting with an infected.
Parameter $\gamma$ is the rate an infected 
node recovers and moves into the resistant 
phase. 
To simulate infection spreading, the model 
is started with a set of seed nodes and 
expecting other nodes to copy the attributes 
or the behaviours of the seed nodes.

Label propagation model~\cite{%
	raghavan2007near} 
runs over a society with a particular group 
carrying an attribute, called \emph{label}. 
In each social interaction, 
individuals copy the label with the highest frequency. 
At the end, each node has a particular label. 
When the copying behaviour is reversed, e.g., the 
attribute is not copied but pushed to the 
individual forcefully, the term ``spread'' springs. 
Rumor spreading model~\cite{%
	daley1965stochastic}  
roots to infection spreading model~\cite{%
	anderson1992infectious}.
The model proposes three types of individuals: 
spreaders, who are willingly spread the rumor in the network; 
ignorant, who does not care about the rumor, 
and 
the stifflers who deliberatively stops the spreading. 

\begin{table}[H]
	\centering
	\caption{
		Network metrics of the datasets. 
	}
	\begin{tabular}{|l|r|r|r|r|r|r|}
		\hline
		\textbf{Dataset} & \multicolumn{1}{c|}{\textbf{$\hbAbs{V}$}} & \multicolumn{1}{c|}{\textbf{$\hbAbs{E}$}} 
		& \multicolumn{1}{c|}{\textbf{\# shells}} & \multicolumn{1}{c|}{\textbf{\# core}}   \\ \hline
		Epinions~\cite{%
		richardson2003trust}     &  75,879 &   508,837 &   67  & 486 \\ \hline
		Email-Eu-core~\cite{%
		leskovec2007graph,%
		yin2017local}            &   1,005 &    25,571 &   34  &  79 \\ \hline
		Wiki-Vote~\cite{%
		leskovec2010signed}      &   7,115 &   103,689 &   53  & 336 \\ \hline
		CA-Cond-Mat~\cite{%
		leskovec2007graph}       &  23,133 &    93,497 &   25  &  26 \\ \hline
		DBLP~\cite{%
		yang2015defining}        & 317,080 & 1,049,866 &   113 & 114 \\ \hline
	\end{tabular}
	\label{table:network-metrics}
	
	$\hbAbs{V}$ and $\hbAbs{E}$ is the total number of nodes and edges. 
	\# shells is the number of shells found with $k$-shell decomposition. 
	\# core is the population of the core.

\end{table}

\subsection{Data Set}

We test our approach on real-life networks~\cite{%
	richardson2003trust,%
	yin2017local,%
	leskovec2010signed,%
	leskovec2007graph,%
	yang2015defining%
}. 
We used 
a trust network~\cite{%
	richardson2003trust},
an e-mail network~\cite{%
	leskovec2007graph,%
	yin2017local},
a voting network~\cite{%
	leskovec2010signed},
and two co-authorship networks~\cite{%
	leskovec2007graph,%
	yang2015defining}.  
The sizes of the networks vary between 
 $1,000$ nodes to 
 $300,000$ nodes.
See \reftbl{table:network-metrics} for network metrics.

\begin{itemize}

	\item 
	\textbf{Epinions.} 
	This is a who-trust-whom online social network of 
	a general consumer review site Epinions.com~\cite{%
		richardson2003trust}. 
	Members of the site can decide 
	whether to trust each other. 
	If a member $i$ trusts the member $j$,
	the graph contains an edge between $i$ to $j$.

	\item 
	\textbf{Email-Eu-core.} 
	This is an anonymized directed network generated using e-mail data 
	from a large European research institution~\cite{%
		leskovec2007graph,%
		yin2017local}. 

	Let $i$ and $j$ be two nodes in the network. 
	There is an edge $(i, j)$ in the network 
	if person $i$ sent person $j$ at least one e-mail. 
	The e-mails only represent communication between institution members, 
	and the dataset does not contain incoming messages 
	from or outgoing messages to the rest of the world.

	\item 
	\textbf{Wiki-Vote.} 
	This directed network contains all the Wikipedia voting data 
	for choosing administrators from the inception of Wikipedia 
	till January 2008~\cite{%
		leskovec2010signed}. 
	Nodes in the network represent wikipedia 	users 
	and a directed edge from node $i$ to node $j$ 
	represents that user $i$ voted on user $j$.

	\item 
	\textbf{CA-Cond-Mat.} 
	This is the Condense Matter Physics collaboration network
	from arXiv.org~\cite{%
		leskovec2007graph}. 
	If an author $i$ co-authored a paper with author $j$, 
	the graph contains an undirected edge from $i$ to $j$. 

	If the paper is co-authored by $n$ authors, 
	this generates a clique of $n$ nodes.

	\item 
	\textbf{DBLP.} The DBLP  provides 
	a comprehensive list of research papers in computer science~\cite{%
		yang2015defining}. 
	The network contains co-authorship relationships,
	where two authors are connected 
	if they publish at least one paper together. 

\end{itemize}

\section{Proposed Method}

We speculate that the approach of 
Kitsak et al.~\cite{%
	kitsak2010identification}
suffers the problem of using seeds that are 
too close to each other. 
That is, if two agents are too close, 
using both as seeds reduces the potential 
reach capability. 

It is better to use one of them as seed 
and select another agent that is not ``close'' 
to the selected one. 
Delivering all the seed notes among the members 
of a particular shell performs weakly in terms 
of exposing the information to other shells, 
since they will spread the idea to other informed 
nodes, not the uninformed nodes. 
Additionally, since the innermost shell 
of a real-life network contains incomparably 
more number of nodes than a practically 
feasible number of seeds that we want to 
choose, we need to find a smart way to 
distribute the information we have.

We propose a method for finding a set of 
influencers who can maximize the spread 
of the information in the network, which 
performs better than degree centrality, 
eigenvector centrality and $k$-shell 
decomposition itself.
To achieve this, we decompose network using 
$k$-shell decomposition, then we pick a set of 
influencers in each shell and the size of the 
selection for each shell is proportional to 
the shell's population. 
Additionally, we modified SIR model and 
introduce hyper-short infected state, 
and propose Uninformed/Informed model.

\subsection{Combining $k$-shell Number 
Approach with Communities}

The innermost shell includes a highly 
connected group of nodes. 
Infecting a subset of the innermost 
shell may be enough to spread the information 
to the all nodes in the innermost shell. 
Instead of selecting all nodes in the 
innermost shell as seed nodes, we select 
a subset of them, then we select other nodes 
from different shells to increase the spreading.

We propose two methods, 
which focus not only the core but also to other shells  
when choosing the seed nodes.

\begin{itemize}
	\item 
	\textbf{$k$-shell proportional (\algP).} 
	$n$ seed nodes are chosen among different 
	shells, proportional to shell population. 
	When choosing nodes within a shell, we 
	sort the nodes by their degree and select 
	the top nodes.
	Assume shell populations are 100, 60, 40 
	from outer shell to inner shell. 
	Proportional seed selection from these shells 
	will result 5, 3, 2 seeds from these shells 
	respectively.
	If number of shells is greater than number of 
	seeds, then we share seeds starting from 
	the core to outer shells proportional to 
	shell population, until we run out of seeds.

	\item 
	\textbf{$k$-shell half proportional (\algH).} 
	Nodes are sorted using their $k$-shell index. 
	First $n/2$ nodes are chosen among the most inner shell. 
	If there are more 
	than $n/2$ nodes in the most inner shell, we sort the nodes by their degree
	and get top $n/2$ nodes.
	Remaining $n/2$ seeds are chosen using $k$-shell proportional method. 
	
\end{itemize}

\begin{figure}[H]
	\centering
	\includegraphics[width=\columnwidth]
		{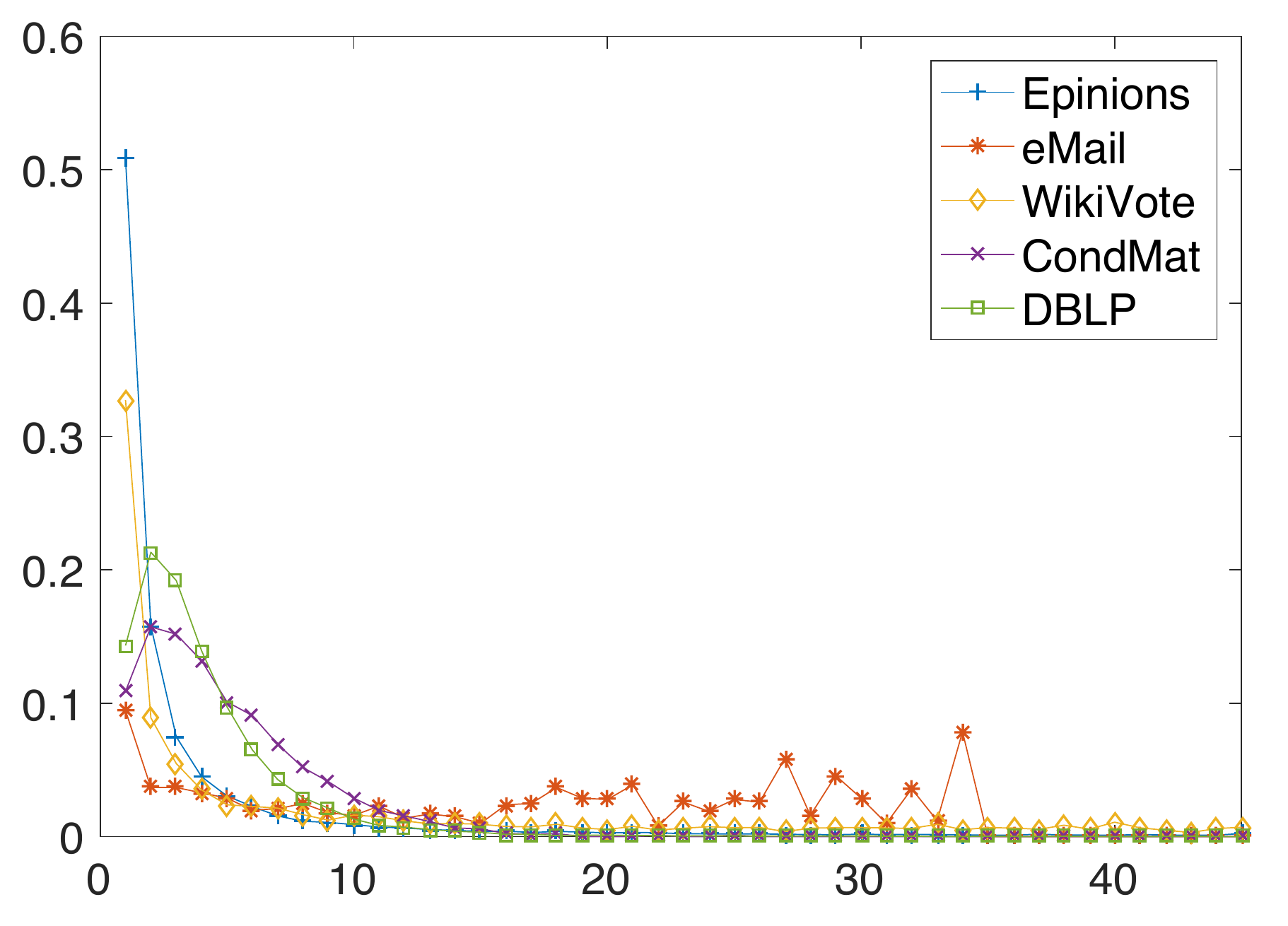}
	\caption{
	$k$-shell index population 
    distribution of \datM, \datE, \datW,
    \datD\ and \datC\ datasets.
	}
	\label{fig:core-distributions}%
	\end{figure}

It is expected that the population decreases as 
one goes to the inner shells.
The populations distributions of the shells are 
given in \reffig{fig:core-distributions} 
for the four networks confirm this.
(The first 40 shells are shown only, since the
remaining shells are characteristically similar.)
But the e-mail data set, is an exception.
It behaves similarly in the first 15 shells,
but then it follows a different pattern.
Populations stop decreasing.
This may be due to the fact that e-mail dataset 
is pruned on creation. 
It represents an e-mail network among 42 
departments in an EU institution 
and e-mail sent out of the institution were pruned.	


\section{Method of Comparison}

We use slightly modified infection spreading to 
compare algorithms for influencer detection. 
Given number of seed nodes $n$,
we run two algorithm to identify two sets of 
$n$ influencers.
Then we run infection spreading algorithm on 
the network, where initially the selected $n$ 
influencers are in infected state 
and the rest of the node in susceptible state.
We conclude that the influencer detection 
algorithm that results higher number of 
infected nodes is better.

\subsection{Uninformed/Informed Model}
We modify SIR model of infection spreading 
to spreading of ideas/opinions. 
It is clear that once an idea is 
introduced to a node, there is no way 
that the node forget the idea. 
Therefore 
The model does not have ``recovered'' state. 
Hence, 
we the model has two states: 
(i)~\emph{uninformed} (U) nodes 
that are not exposed to the idea 
and
(ii)~\emph{informed} (I) nodes 
that are already exposed to the idea.

In our era, an individual is exposed to 
too many ideas every hour, so that they
spread the ideas for a very brief time period,
to their close acquaintances then stops spreading.
Hence in our model, 
we assume that a node, 
that is just exposed to the idea, 
has very short time to share that to its 1-hop
neighbors.
After that time, they stop sharing. 

In this sense,
we have ``active'' informed state 
that the node is actively sharing 
and ``inactive'' informed state, 
which the node does not share. 
With this configuration, 
we start the simulation with a set of ``active informed''  nodes 
and the rest of the network is inuninformed state. 
Representing a social-interaction in real-life,
if an active node interacts with an uninformed node,
the uninformed node,
as in the case of SIR model, 
gets the idea with probability of $\beta$
and becomes an active informed.
Hence ready to propagate the idea even further.
When $\beta$ is chosen too large, the information
spreads to entire network. 
When it is chosen too small, no matter how to 
choose the spreaders, the information cannot
be spread over the network at all.
To compare the methods clearly, we use $\beta = 0.09$,
which is a typical value in this 
domain~\cite{%
	zhang2016identifying}.
Notice that we do not need parameter $\gamma$.

\subsection{Comparison}

We set the nodes, 
that are indicated as $n$ influencers by an influencer detection algorithm,
to informed state.
Then let the idea propagate in the network by simulating the uninformed/informed model.
We give enough time to simulation to converge,
i.e.,
no further propagation occurs.
We define the performance of an influencer detection algorithm 
by the percentage of the nodes that are in the informed state at the end of the simulation.
Clearly, this is a stochastic process, 
therefore the average of $100$ realizations is reported.
An influencer detection algorithm with the highest percentage of propagation is the best one.

\section{Experiments}

We run two sets of experiments. 
In the first set, 
we picked a fixed number of seed nodes independent of the network size. 
In the second set, 
we set the number of seed nodes proportional to the size of the network. 
In both sets of experiments,
we compare
Degree (Dg),
Eigenvector Centrality (Eg),
PageRank (Pr),
$k$-shell ($k$-shell)
methods with our proposed methods,
namely,
$k$-shell proportional (Ks-P) and
$k$-shell half proportional (Ks-Hp).

\textbf{Experiment with fixed number of seeds.}
In many promotional campaigns, 
the budget has a well-defined limit 
such as 
a fixed amount of money or 
a number of promotional items. 
To mimic such cases,
we assumed that there are a fixed number 
$n$ of seed nodes,
which is independent of the target size, 
i.e., the number of nodes.
We choose $n = 100$, 
which happens to be $10\%$ of the size of 
the smallest network in our dataset.

\begin{figure}[H]
	\centering
	\includegraphics[width=\columnwidth]
		{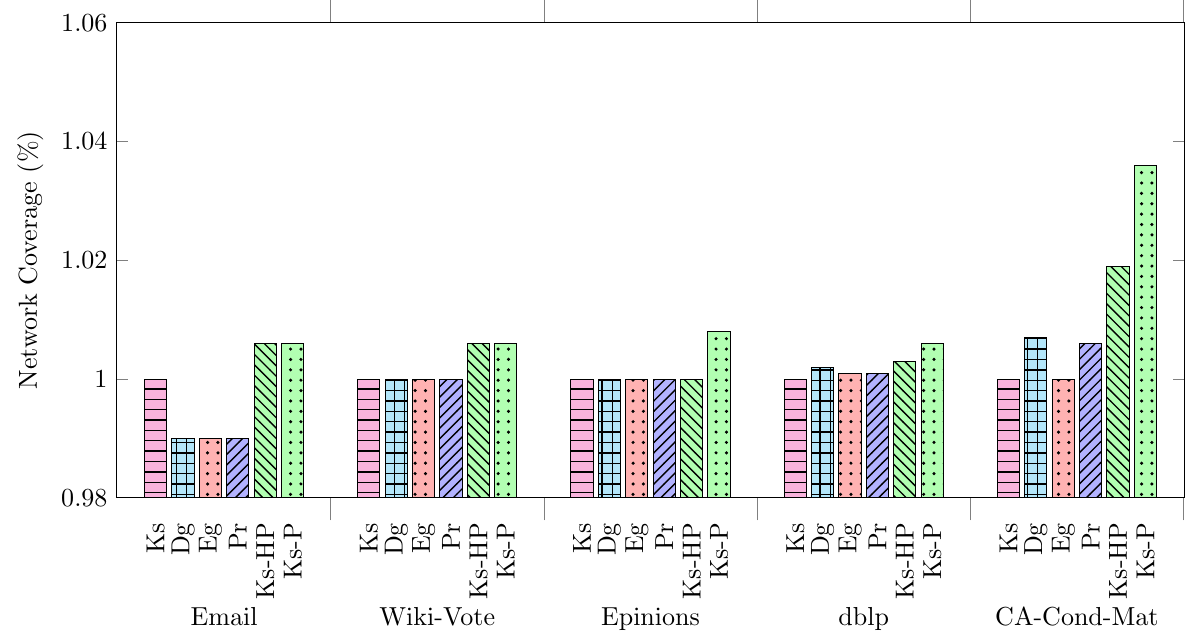}
	\caption{
		Experiment-1.
		The number of seeds is $100$ independent 
		of the network size.
		(The results are averaged over 
		\textbf{100} independent runs)
	}
	\label{fig:experiment-results-1}%
\end{figure}

\textbf{Experiment with proportional number 
of seeds.}
A more reasonable scenario is to select $n$ 
proportional to the network size.
For this case, we arbitrarily select $10\%$ 
of the network size, i.e., $n = N / 10$.
This enables us to compare results of the 
two experiments.
Note that in the first experiment 
we used $10\%$ of our smallest network, 
e-mail network, as the seed.
So the results of Email network in both figures 
are the same.

\begin{figure}[H]
	\centering
	\includegraphics[width=\columnwidth]
		{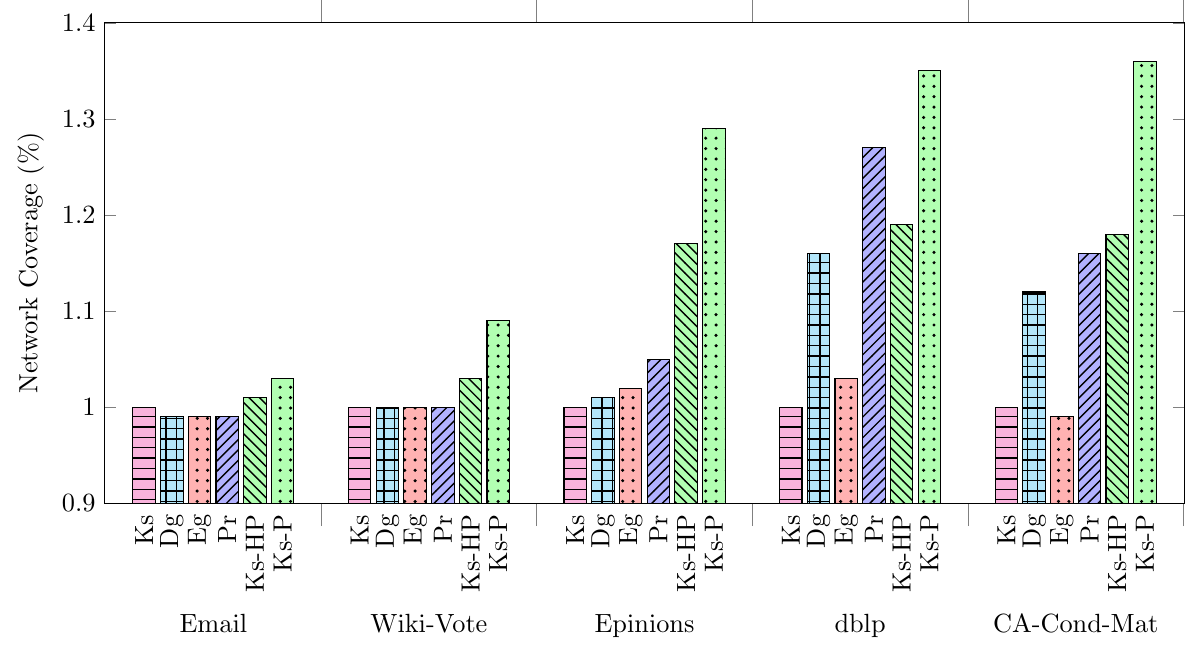}
	\caption{
		Experiment-2.
		The number of seeds is $10\%$ of the 
		network size.
		(The results are averaged over \textbf{100} 
		independent runs)
	}%
	\label{fig:experiment-results-2}%
\end{figure}

\subsection{Discussion}

We compare each algorithm by running them on our 
five data sets and observing their network coverages.
One could provide the percentage of coverages 
but the results are quite close to each other, 
especially in the small networks such as Email 
and Wiki-Vote networks.
So we prefer another presentation.
Since our proposed algorithms are based on 
$k$-shell algorithm, we use the network coverage 
of $k$-shell algorithm as the comparison unit.
The coverages of other algorithms are presented 
as relative to that of $k$-shell,
which we call it \emph{relative coverage}.
Clearly the relative coverage of $k$-shell would 
be 1 in this presentation.
Any algorithm with better coverage than that 
of $k$-shell would have a relative coverage 
that is larger than 1.
In \reffig{fig:experiment-results-1} and 
\reffig{fig:experiment-results-2}
relative coverages of the algorithms are 
grouped for each network for ease of comparison.

The first set of experiments uses 100 seed 
nodes which is clearly insufficient for the 
spreading of the ideas in larger networks. 
In \reffig{fig:experiment-results-1} we 
observe that all algorithms perform identical 
for \datM\ and \datW.

Our algorithm, \algP\, performs slightly 
better ($1-4\%$) in \datE\, \datD\ and \datC. 
Although \algP\ is slightly better than other 
algorithms given a fixed number of seed nodes, 
it can be clearly seen larger networks requires 
larger set of seeds.

The results of the second experiment are given 
in \reffig{fig:experiment-results-2}.

Our algorithms are slightly better performed.
Interestingly, while for the other networks 
degree and eigenvector centrality outperform 
$k$-shell, they perform poorly compared to 
$k$-shell in this network.

This may be because of the pruning of the 
e-mails that goes to outside of the organization.

For Wiki-Vote network, 
which is the second smallest network,
algorithms degree, eigenvector centrality 
and $k$-shell have so similar performances 
that even our presentations cannot visualized 
the differences.
Our algorithms are clearly better performed 
in this network.
Especially our proportional algorithm, Ks-P, 
has almost $10\%$ better relative coverage.
That is, around $368$ more nodes in a network 
of $7,115$ nodes.

Starting \datE\ network,
\algD\ and \algE\ outperform \algK.
Yet our algorithms are clearly much better.
\algH\ gets almost $20\%$, and \algP\ close 
to $30\%$ more relative coverage.
That means $3,467$ and $5,977$ 
more nodes in a network of 75,879.

In \datD\ network,
all algorithms perform better than \algK\
while \algP\ manages to reach to relative 
coverage of more than $34\%$.
That means an impressive additional $25,737$ 
more nodes in a network of $317,080$ nodes.
Note that this time \algD\ becomes the fourth 
after \algX\ and \algH.

In \datC\ network \algP\ performs its best with 
a relative coverage of $36\%$.
Relative to \algK\ \algP\ can reach to $2,409$ 
more nodes in  a network of $23.133$
nodes.

The results indicates that, our method, Ks-P 
outperforms existing methods by 3-36\%.

\section{Conclusion}

Since word-of-mouth diffusion is an important 
model for exposing and information into a 
network of individuals, finding a set of 
influencers in the network becomes an 
important task in viral marketing. 
In this study, we proposed an extension 
to \algK\ decomposition method for maximizing 
the network coverage with the same size of 
set of influencers. 
Our experiments demonstrate that our method 
outperforms degree centrality, eigenvector 
centrality and \algK\ method, which is 
up to 36\% better than
the mentioned algorithms in different datasets.

\section*{Acknowledgment}
We used NepidemiX Python Package to implement 
the model \cite{%
	Nepidemix}. 
\ifCLASSOPTIONcaptionsoff
  \newpage
\fi

\IEEEtriggeratref{32}
\bibliographystyle{IEEEtran}
\bibliography{IEEEabrv,k-shell-extended-main-tex}

\end{document}